\documentclass[11pt,a4paper]{article}
\RequirePackage{amsmath,amssymb}
\RequirePackage[dvipsnames,usenames]{color}

\usepackage{cite}
\usepackage{fullpage}

\usepackage[british]{babel}
\usepackage[latin1]{inputenc}
\usepackage[T1]{fontenc}
\usepackage[final]{showkeys} 
\usepackage[bookmarks]{hyperref}
\usepackage{amsthm}
\usepackage{graphicx}
\usepackage{subfigure}
\usepackage{braket}
\usepackage{mathrsfs}

\usepackage{bm}
\usepackage{amsfonts}
\usepackage{braket}
\usepackage{graphics}

\newcommand{\bc}{\begin{center}}
\newcommand{\ec}{\end{center}}

\newcommand{\be}{\begin{equation}}
\newcommand{\ee}{\end{equation}}

\title{\bf Emmy's letter to Santa Claus (and a reply):\\ 
Vaidya geometries and scalar fields with null gradients}

\author{
Valerio~Faraoni, Andrea Giusti, and Bardia H. Fahim
$\,$
\\
\\
Department of Physics \& Astronomy, Bishop's University
\\
2600 College Street, Sherbrooke Qu\'ebec, Canada J1M~1Z7
}

\begin{document}
\maketitle
\begin{abstract}

Emmy, a child prodigy, writes a letter to Santa Claus asking for a Vaidya 
solution sourced by a massless scalar field with null gradient for 
Christmas. Santa writes back explaining why such a toy cannot be built and 
offers an exact plane wave sourced by a scalar field with lightlike 
gradient as a replacement.

\end{abstract}

\newpage

\section{Emmy's letter to Santa Claus}
\label{sec:1}
\setcounter{equation}{0}

Dear Santa,\\ Can I please have a Vaidya solution of the Einstein 
equations $G_{ab}=8\pi T_{ab}$ sourced by a scalar field $\phi$ with null 
gradient this Christmas?  I really would like to play with it. My friends 
in string theory and holography class play with Vaidya toys 
\cite{Myers,Bhatta}. Being just a first grader, I play with simple scalar 
fields instead, but I like null vectors, null dust sprinkled all over 
spacetime, and flashy surfaces moving at the speed of light.

As you certainly remember, an outgoing Vaidya solution 
\cite{Vaidya1,Vaidya2} (see also \cite{Stephani,Poisson}) is given by the 
line element\footnote{We follow the notation of Ref.~\cite{Waldbook} and 
use units in which the speed of light and Newton's constant are unity.}
\be
ds^2=-\left( 1-\frac{2m(u)}{r} \right) du^2 -2dudr +r^2 d\Omega_{(2)}^2 
\,,\label{outgoingVaidya}
\ee 
where $u$ is a retarded null coordinate, $m(u)$ is a regular mass 
function, and 
$d\Omega_{(2)}^2 \equiv d\vartheta^2 +\sin^2 \vartheta \, d\varphi^2$ is 
the line 
element on the unit 2-sphere.  The only non-vanishing component of the 
Ricci tensor is
\be
R_{uu} = -\, \frac{2m'(u)}{r^2} \,,
\ee
where a prime denotes differentiation with respect to $u$. The 
stress-energy tensor of the null dust sourcing the Vaidya geometry is 
\be
T_{ab}= -\, \frac{m'(u)}{4\pi r^2} \, l_a l_b \,,
\ee
where the vector $\ell^a $ is null and must be geodesic, as follows from the 
conservation equation for the null dust $\nabla^b T_{ab}=0$ (see 
\cite{Kuchar} for a review on the null dust). For the ougoing Vaidya 
metric 
it is $\ell_a=-\partial_a u$ \cite{Poisson,Stephani}.

A null dust appears frequently in classical and quantum gravity, in Vaidya 
spacetimes \cite{Vaidya1,Vaidya2}, $pp$-waves 
\cite{Kuchar,Griffiths,Krasinski}, 
Robinson-Trautman geometries \cite{RobinsonTrautman}, twisting 
solutions of the Einstein-Maxwell equations 
\cite{RobinsonTrautman,Herlt,twisting2}, in studies of classical and 
quantum gravitational collapse, horizon formation, mass inflation  
\cite{collapse1,collapse2, collapse3, collapse4, collapse5, collapse6, 
collapse7, collapse8, collapse9, collapse10}, black hole evaporation 
\cite{evaporation1,evaporation2, evaporation3}, and canonical Hamiltonians 
\cite{Kuchar,Horvath, Myers}.

I really would like this null dust to be realized by an outgoing scalar 
field  $\phi $ with null gradient, $\nabla_c \phi \nabla^c \phi=  0$, say 
$\phi=\phi(u)$.  I would also be happy with an ingoing Vaidya solution of 
the form
\be
ds^2=-\left( 1-\frac{2m(v)}{r} \right) dv^2 +2dvdr +r^2 d\Omega_{(2)}^2 
\,, \label{ingoingVaidya}
\ee
where $v$ is the advanced null coordinate, with  $\phi(v)$ going with it.

Can you please bring any of these two toys this Christmas? Many 
(advanced) thanks, \\\\
Emmy

\section{Santa's reply}
\label{sec:2}
\setcounter{equation}{0}

Dear Emmy,\\\\ 
Sorry for this retarded reply. Unfortunately, the toy that you are asking 
for cannot be built in my workshop or anywhere else in this universe. Here 
is why. First, we need to talk about the null dust sprinkled all over 
Vaidya's spacetime and whether a scalar field can be equivalent to a null 
dust.

It is well known \cite{Ellis71,Madsen88,Faraoni12} that a scalar field 
$\phi$ with timelike gradient, $\nabla^c\phi\nabla_c \phi<0$ is equivalent 
to an irrotational perfect fluid with four-velocity
\be
u_c= \frac{ \nabla_c\phi}{ \sqrt{ -\nabla^a\phi\nabla_a\phi}} \,.
\ee
Scalar fields with spacelike gradients 
are less interesting and seldom considered from this point of view 
\cite{Semiz}. But what if $\nabla^c\phi \nabla_c \phi=0$ instead? Well, 
the stress-energy 
tensor of a massless, minimally coupled, scalar field $\phi$ is simply 
$T_{ab}=\nabla_a \phi \nabla_b\phi -\frac{1}{2} \, g_{ab} \nabla^c \phi 
\nabla_c \phi $, which matches that of  a null dust if
$\ell_a \equiv \nabla_a \phi$ is null. Because $\ell^a$ is a gradient, 
only an irrotational dust can be reproduced this way.  Now, the divergence 
of $\ell^a$ is 
\be
\nabla^c \ell_c = \nabla^c \nabla_c \phi \equiv \Box \phi =0
\ee
by virtue of the Klein-Gordon equation of motion for $\phi$.  The 
covariant conservation equation $\nabla^b T_{ab}=0$ satisfied by any 
fluid, including the null dust, gives
\be
\ell^b \nabla_b \ell_c = -\left( \nabla^a \ell_a \right) \ell_c 
\ee
which is the non-affinely parametrized geodesic equation. However, since 
$\Box\phi=0$, the right hand side vanishes and  $\ell^a$ must be 
null and tangent to affinely-parametrized null geodesics.  

There is a subtlety here: since a null vector is defined up to 
a multiplicative (positive) factor, one could choose a different 
parametrization of the null dust in which its stress-energy tensor is 
instead written as $T_{ab}=\rho \, \ell_a \ell_b$ \cite{Kuchar,ValerioJeremy}. For 
a general null dust, with this choice the covariant conservation equation 
$\nabla^b 
T_{ab}=0$ would then give non-affinely parametrized geodesics 
\cite{Kuchar}. However, because the scalar field null dust is 
irrotational, also this parametrization ends up giving affine 
parametrization \cite{Kuchar,ValerioJeremy}.

Now consider the congruence of null geodesics tangent to the null vector 
$\ell^a$ (outgoing null geodesics for the geometry~(\ref{outgoingVaidya}), or 
ingoing null geodesics for the geometry~(\ref{ingoingVaidya})). Let us 
focus on the outgoing Vaidya spacetime~(\ref{outgoingVaidya}) first: by 
comparing the relation $\ell_a=-\nabla_a u  $ valid in Vaidya's geometry 
with $\ell_a=\nabla_a \phi$ specifying that the null dust is generated by a 
scalar field, 
we obtain (apart from an irrelevant additive integration 
constant) $\phi=-u$, so the scalar field is outgoing at the speed of 
light. 
However, this result is in contradiction with the Klein-Gordon equation 
$\Box \phi=0$ because
\begin{eqnarray}
\Box\phi &=& \nabla^c \ell_c = - \nabla^c \nabla_c u = -\frac{1}{\sqrt{-g}} 
\,  \partial_{\mu} \left( \sqrt{-g} \, g^{\mu\nu} \partial_{\nu} u \right) 
\nonumber\\
&&\nonumber\\
&=& - \frac{1}{\sqrt{-g}} \, \partial_{\mu} \left(  \sqrt{-g} \, g^{\mu u} 
\right) = - \frac{1}{ r^2 \sin\vartheta } \, \partial_r \left( r^2 
\sin\vartheta \, 
g^{ur}   \right) = \frac{2}{r} \,,
\label{questa}
\end{eqnarray}
where we used the fact that 
\be
\left( g^{\mu\nu} \right) = \left( \begin{array}{cccc}
0 & -1 & 0 & 0 \\
&&&\\
-1 & 1-\frac{2m}{r} & 0 & 0 \\ 
&&&\\ 
0 & 0 & \frac{1}{r^2} & 0 \\ 
&&&\\ 
0 & 0 & 0 & \frac{1}{r^2 \sin\vartheta} \\ 
\end{array} \right) 
\ee 
and $\sqrt{-g}=r^2 \sin\vartheta$.  Clearly, $\Box \phi = -2/r \neq 0$ and 
the Klein-Gordon equation cannot be satisfied. Although a scalar 
field with lightlike gradient is a null dust realization, a  
scalar field-sourced outgoing Vaidya geometry is impossible. More 
generally, if one considers radial outgoing and ingoing null geodesics 
with tangents $\ell^a$ and $n^a$ in the outgoing Vaidya 
geometry~(\ref{outgoingVaidya}), their expansions are easily found to be 
({\em e.g.}, \cite{Poisson})
\be 
\theta_{(\ell)}= \nabla_c \ell^c = \frac{2}{r} 
 \,, \;\;\;\;\;\;\;\;\;\;\; \theta_{(n)}= \frac{2m(u)-r}{r^2} \,. 
\ee 
Clearly, both expansions are non-vanishing for $r>2m$. Put in other words, 
the 
congruence of outgoing radial null geodesics that is deemed to have 
$\ell^c=\nabla^c \phi$ as its four-tangent field is twist-free (because $\ell^c 
$ is a gradient), shear-free (because of spherical symmetry), and cannot 
also be expansion-free.

The same argument, adapted to the replacement $u \rightarrow v$ 
demonstrates the impossibility of ingoing Vaidya solutions sourced by a 
scalar field (that would have to be $\phi=-v$). So, it is impossible to 
build your toy! But don't be disheartened: since you like null dust and 
surfaces travelling at light speed I can offer you, as a replacement, 
an exact plane wave sourced by a scalar field with null four-gradient. 
This plane wave, described in \cite{Bressange}, is given in null 
coordinates by the Szekeres form \cite{Szekeres} 
\be
ds^2 =-2\, \mbox{e}^{-M(u)} dudv+\mbox{e}^{-U(u)} \left( \mbox{e}^{V(u)} 
dx^2 + \mbox{e}^{-V(u)} dy^2 \right) \,,
\ee  
where \cite{Bressange} 
\begin{eqnarray}
\mbox{e}^{-U} &=& f(u)+\frac{1}{2} \,,\\
&&\nonumber\\
\mbox{e}^{V} &=& \left( \frac{ 1+ \sqrt{ \frac{1}{2} -f}}{1-\sqrt{ 
\frac{1}{2}-f} } \right)^{\lambda_1/2} \,,\\ 
&&\nonumber\\
\mbox{e}^{-M} &=& kf'(u) \, \frac{ \left( \frac{1}{2}+f \right)^{ 
\frac{\alpha^2-1}{2} } }{ \left( \frac{1}{2}-f \right)^{\alpha^2} } 
 \,,\\
&&\nonumber\\
\phi &=& \phi_0 +\frac{\lambda_2}{2} 
\ln \left( \frac{ 1+\sqrt{ \frac{1}{2}-f} }{ 1-\sqrt{ \frac{1}{2}-f} } 
\right)\,,
\end{eqnarray}  
and where $f(u)$ is an arbitrary (but  regular) function of $u$, 
$\lambda_{1,2}$ and $k$ are constants, and \cite{Bressange} 
\be
\alpha^2 = \frac{\lambda_1^2}{4}+\lambda_2^2 \,.
\ee
In this case we have
\be
\nabla_{\mu} \phi = -\frac{\lambda_2 f'(u) }{\sqrt{\frac{1}{2}-f} \left(
1+ 2 \, f \right)} \, \delta_{\mu u}
\ee
and\\
\be
\left( g^{\mu\nu} \right)=\left( \begin{array}{cccc}
0 & -\mbox{e}^{M} & 0 & 0 \\
&&&\\
-\mbox{e}^{M} & 0 & 0 &0 \\
&&&\\
0 & 0& \mbox{e}^{U-V} & 0  \\
&&&\\
0 & 0& 0 & \mbox{e}^{U+V}\\
\end{array} \right) \,,
\ee
using which one obtains
\be
\nabla^c \phi \nabla_c \phi = g^{uu}\phi'(u)=0 \,.
\ee 
In this case $\ell^c \equiv \nabla^c \phi$ can be a null vector without 
contradicting the vanishing of the divergence $\nabla^c \ell_c= \Box \phi= 
0$ because we have planar, instead of spherical, symmetry and null 
geodesics with tangents $\ell^c$ can form a non-expanding congruence.\\\\
Yours,\\
Santa

\section{Epilogue}
\label{sec:3}
\setcounter{equation}{0}

Santa has demonstrated that Vaidya geometries sourced by a massless scalar 
field acting as a null dust cannot be solutions of the Einstein equations. 
The physical reason is that these spherically symmetric geometries must 
necessarily have non-vanishing expansion, which contradicts the 
Klein-Gordon equation. In fact, an (irrotational) null dust created from a 
lighlike scalar field gradient $\ell_a=\nabla_a\phi$ must necessarily have 
zero 
divergence $\nabla^c \ell_c=\Box\phi$, which is incompatible with the 
expanding (or contracting) nature of the Vaidya geometries.

Emmy, who likes to play with symmetries, is happy anyway about the new 
plane wave toy that she will receive this Christmas (which can be built 
because, contrary to Vaidya, it can simultaneously have zero expansion and 
satify the Klein-Gordon equation). Along the way, the child has also 
learned something interesting from her correspondence with Santa.

{\small 
\section*{Acknowledgments}

This work is supported, in part, by the Natural Sciences \& Engineering 
Research Council of Canada (Grant No.~2016-03803 to V.F.) and by Bishop's 
University. The work of A.G. has been carried out in the framework of the 
activities of the Italian National Group for Mathematical Physics [Gruppo 
Nazionale per la Fisica Matematica (GNFM), Istituto Nazionale di Alta 
Matematica (INdAM)].

\end{document}